\documentclass[preprint,prb,superscriptaddress,amssymb,amsmath, floatfix,citeautoscript]{revtex4}

\usepackage[pdftex]{graphicx}

\newcommand{\p}{\partial }
\newcommand{\w}{\omega}
\renewcommand{\k}{\mathbf{k}}

\newcommand{\Tzt}{T(z,t)}
\newcommand{\vzk}{v_{z\k}}
\newcommand{\vyk}{v_{y\k}}

\newcommand{\tk}{\tau_{\k}}
\newcommand{\lzk}{\Lambda_{z\k}}
\newcommand{\lyk}{\Lambda_{y\k}}

\begin{document}

\title{Thermal phonon boundary scattering in anisotropic thin films}

\author{A. J. Minnich}
\email{aminnich@caltech.edu}
\affiliation{Division of Engineering and Applied Science\\
California Institute of Technology\\
Pasadena, CA 91125}

\begin{abstract}
Boundary scattering of thermal phonons in thin solid films is typically analyzed using Fuchs-Sondheimer theory, which provides a simple equation to calculate the reduction of thermal conductivity as a function of the film thickness. However, this widely-used equation is not applicable to highly anisotropic solids like graphite because it assumes the phonon dispersion is isotropic. Here, we derive a generalization of the Fuchs-Sondheimer equation for solids with arbitrary dispersion relations and examine its predictions for graphite. We find that the isotropic equation vastly overestimates the boundary scattering that occurs in thin graphite films due to the highly anisotropic group velocity, and that graphite can maintain its high in-plane thermal conductivity even in thin films with thicknesses as small as ten nanometers.

\end{abstract}

\maketitle

Thermal transport in thin solid films with thicknesses from tens of nanometers to micrometers is a topic of considerable importance,\cite{Cahill2014,Ruan2014,XinjingWang2015,Balandin2012} with such films being used in applications ranging from quantum well lasers to LED lighting.\cite{Chowdhury:2009,Su:2012a, Moore2014} In these thin films, the phonon mean free paths (MFPs) are comparable to the film thickness, resulting in boundary scattering that reduces the thermal conductivity and inhibits heat dissipation. This thermal management problem is presently an important challenge in many devices such as GaN transistors.\cite{Balandin2011, Su:2012a,Goodson2014} 

Heat transport along the in-plane direction of thin films is typically described using the Fuchs-Sondheimer equation, which is an analytical solution of the Boltzmann transport equation (BTE) for thin films with partially specular and partially diffuse walls. This equation was originally derived for electron transport \cite{Fuchs1938, Sondheimer1952} and later extended to phonon thermal transport assuming an average phonon MFP, enabling the calculation of thermal conductivity as a function of the film thickness.\cite{Chen1993,Chen1997} Mazumder and Majumdar used a Monte-Carlo method to study the phonon transport along a silicon thin film including phonon dispersion and polarizations.\cite{Mazumder2001}

While the Fuchs-Sondheimer equation is in wide use, it cannot be applied to anisotropic transport in its typical form because it assumes the crystal of interest is isotropic. However, there are many situations in which boundary scattering occurs in thin anisotropic films, with the most familiar example being thin graphite films. Such films have been studied experimentally  \cite{jang_thickness-dependent_2010}, and few-layer graphene films have been investigated as in heat spreaders for GaN transistors \cite{Balandin2011}. Provided that the films are sufficiently thick that phonon dispersion modifications in the cross-plane direction can be neglected, mathematically describing thermal transport in anisotropic thin films requires a Fuchs-Sondheimer equation that is valid for any crystal, regardless of its anisotropy. Surprisingly, despite the simplicity of the derivation, no such equation has been reported to the best of our knowledge.

Here, we report the generalization of Fuchs-Sondheimer theory to crystals with arbitrary anisotropies. We find that highly anisotropic solids with small group velocities along certain crystallographic directions experience minimal boundary scattering because the thermal conductivity reduction depends only on the component of the MFP normal to the boundary rather than the overall MFP. As a result, thin films of anisotropic crystals like graphite maintain their high thermal conductivity even as the film thickness becomes very small. This observation has important implications for heat spreading in electronic devices and the thermal conductivity of graphite foams. \cite{pettes_thermal_2012}

We begin by considering the steady BTE in a thin film under the relaxation time approximation. We let the $y$ direction represent the cross-plane direction and assume that a thermal gradient exists along the $z$ direction. Then, the BTE is given by:
\begin{equation} \label{eq:BTE_energy}
 \vyk \frac{\p f_{\k}}{\p y} + \vzk \frac{\p f_{\k}}{\p z} = -\frac{f_{\k}-f_0(\Tzt)}{\tau_{\k}}
\end{equation}
where $f_{\k}$ is the desired distribution function, $\vyk$ and $\vzk$ are the components of the phonon group velocity along the $y$ and $z$ directions, $\tk$ is the phonon relaxation time, and $\k$ is the phonon wavevector in phase space.  This equation is solved for the thin film by letting the small perturbation $g_{\k} = f - f_{0}(\Tzt)$, yielding:

\begin{equation} \label{eq:BTE}
\lyk \frac{\p g_{\k}}{\p y} + g_{\k} = - \lzk \frac{\p f_{0}}{\p T} \frac{\p T}{\p z}
\end{equation}

This result assumes that $\p g_{\k}/\p z \approx 0$ and that the length scale of the thermal gradient is long compared to the phonon MFPs so that a temperature gradient can be defined. Equation \ref{eq:BTE} is a one-dimensional ODE with the boundary conditions $g_{\k}(k_{y}<0, y=d) = g_{\k}(k_{y}>0, y=0) = 0$ corresponding to thermalizing, diffuse scattering, meaning that the phonon distribution emerging from the wall is a thermal distribution at the local boundary temperature. The general solution of this equation is well-known and given in Ref.\ \cite{gangbook} as:

\begin{eqnarray}
g^{+} & = & S_{0}(z) (e^{-\eta} - 1) \\
g^{-} & = & S_{0}(z) (e^{\xi_{y}-\eta} - 1) 
\end{eqnarray}

where $\eta = y/\lyk$, $\xi_{y}=d/\lyk$ is the nondimensional thickness defined by the $y$ component of the MFP, and $S_{0}(z) = - \lzk \frac{\p f_{0}}{\p T} \frac{\p T}{\p z}$. The thermal conductivity can be obtained by substituting this solution into the expression for heat flux:

\begin{equation} \label{}
J_{q} = \int_{0}^{d} \frac{1}{V} \sum_{\k} \hbar \w g_{\k} \vzk dy
\end{equation}

where $V$ is the crystal volume. After evaluating this expression, the thermal conductivity of the thin film along the $z$ direction is identified as:

\begin{equation} \label{kzz}
\kappa_{zz}(d) = \sum_{k_{y}>0} \kappa_{zz, \k} \left(1 - \frac{1 - e^{-\xi_{y}}}{\xi_{y}} \right) =  \sum_{k_{y}>0} \kappa_{zz, i} S(\xi_{y})
\end{equation}

where the intrinsic thermal conductivity $\kappa_{zz,\k} = C_{\k} \vzk^{2} \tk$. $S(\xi_{y})$, which we term the boundary scattering function, is the generalized version of the Fuchs-Sondheimer expression describing the reduction in thermal conductivity that occurs due to boundary scattering. For reference, the traditional Fuchs-Sondheimer expression obtained by integrating over all solid angles in a spherically symmetric Brillouin zone is \cite{Sondheimer1952}:

\begin{equation} \label{kzzorig}
S(\xi) = 1 - \frac{3}{8 \xi} \left[ 1 - 4 E_{3}(\xi) + 4 E_{5}(\xi) \right]
\end{equation}

where $E_{n}(\xi)$ is the exponential integral function and $\xi=d/\Lambda_{\w}$ where $\Lambda_{\w}$ is the isotropic MFP that depends only on phonon frequency.

Comparing Eqs.\ \ref{kzz} and \ref{kzzorig}, we see that the traditional Fuchs-Sondheimer expression averages out the solid angle dependence in phase space so that the final expression depends only on the overall MFP. The generalized expression does not make any assumptions about the symmetry of the crystal, and so the expression depends on the components of the MFPs along the crystal axis normal to the boundary. If this MFP component is different than the value along the in-plane axis, the actual in-plane thermal conductivity obtained with Eq.\ \ref{kzz} may be very different than the prediction of the traditional expression.

To examine this prediction, we consider transport along the basal plane of a thin film of graphite. We use a phonon dispersion calculated from an optimized Tersoff potential provided by Lucas Lindsay \cite{lindsay_optimized_2010}. The actual phonon-phonon relaxation times of graphite are not readily available, and thus we must assume a general form to proceed. In a previous work for which we modeled the graphite dispersion with an anisotropic Debye model,\cite{minnich_phonon_2015} we found that isotropic relaxation times were able to explain the thermal conductivity anisotropy of graphite. In this work for which we use the actual dispersion, we find that these same relaxation times yield $\kappa_{yy}=2000$ W/mK and $\kappa_{zz} =230$ W/mK. Given that the in- and cross-plane thermal conductivities of graphite are 2000 W/mK and 6 W/mK,\cite{touloukian_thermophysical_1970} respectively, this calculation indicates that these relaxation times cannot explain the large thermal anisotropy of graphite.

We now must find the relaxation times that best reproduce the following experimental observations. First, the thermal conductivity along each crystal axis is well-known \cite{touloukian_thermophysical_1970}. Second, recent computational and experimental reports indicate that phonon MFPs along the c-axis are on the order of several hundred nanometers. \cite{fu_experimental_2015,wei_phonon_2014} We satisfy these constraints by assuming relaxation times of the form \cite{pettes_scattering_2015}:

\begin{equation} \label{}
\tau_{p}(\w)^{-1} = A_{p} \left( \frac{\w}{\w_{max,p}} \right)^{2} 
\end{equation}

where $\tau_{p}(\w)$ is the relaxation time for branch $p$, $A_{p}$ is a constant for each branch, $\w$ is the phonon frequency, and $\w_{max,p}$ is the maximum phonon frequency for each branch. Because identifying phonon polarizations off high-symmetry points in the Brillouin zone is challenging, branches are determined by sorting the frequencies obtained from the dynamical matrix from low to high.

We can satisfy the constraints by changing $A_{p}$ for specific branches even though the relaxation times remain isotropic because the different branches contribute differently to thermal conductivity along each crystal axis,  We take $A_{p}=A_{p,0}=8 \times 10^{16}$ s$^{-1}$ for all branches except 1, 2, 3, and 5. Branches 1 and 2 primarily carry heat along the c-axis and must have smaller relaxation times to yield the correct c-axis thermal conductivity. We find that $A_{p,1} = A_{p,0}/100$ and $A_{p,2} = A_{p,0}/20$. Branch 3 must have shorter relaxation times, yielding $A_{p,3} = A_{p,0}/20$. We also cap the maximum relaxation times at 250 ps; otherwise the c-axis MFPs are too long compared to experiment. Branch 5, the LA branch along the ab axis, must have longer relaxation times to reproduce the high ab-axis thermal conductivity, $A_{p,3} = 3 \times A_{p,0}$, with a maximum relaxation time of 500 ps. Finally, the minimum relaxation time of all branches is enforced to be 5 ps so that the MFPs are not unphysically short. These relaxation times yield $\kappa_{ab} = 1500$ W/mK and $\kappa_{c}=6.9$ W/mK, in reasonable agreement with experiment. We emphasize that these relaxation times are not necessarily those of actual graphite, but the thermal transport properties obtained using them are sufficiently close to the experimental values that they can be used for further analysis.


\begin{figure}
\begin{center}
\includegraphics[width=0.8\textwidth]{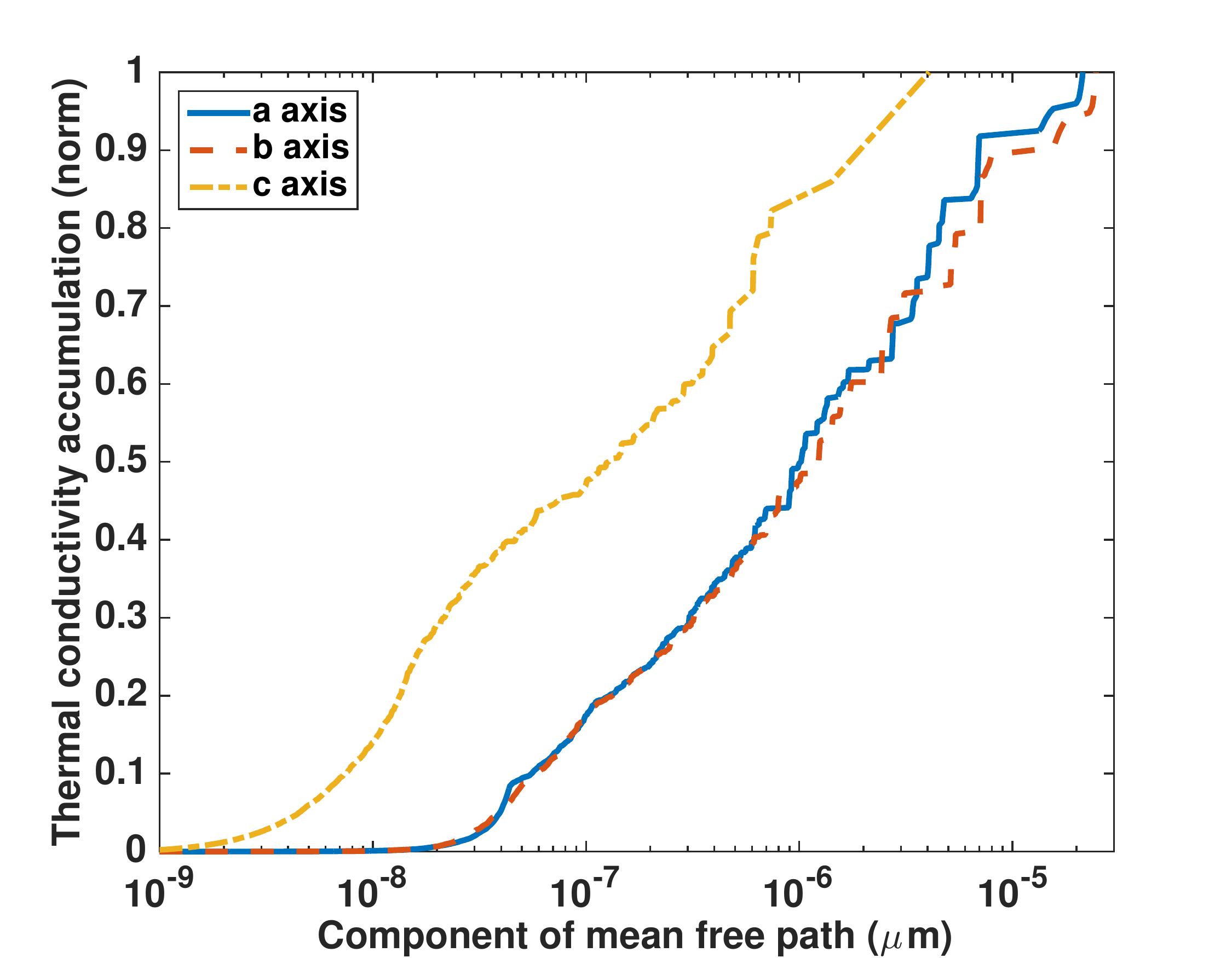}
\caption{Normalized, accumulated thermal conductivity versus component of MFP along the transport direction for each crystal axis using the actual graphite phonon dispersion and semi-empirical relaxation times. The c-axis MFPs are on the order of hundreds of nanometers, in agreement with experiment and considerably shorter than those along the ab-axis.\cite{fu_experimental_2015}}
\label{fig:accum}
\end{center}
\end{figure}

The accumulative thermal conductivity versus the component of the MFP along the a, b, and c crystal axes is shown in Fig.\ \ref{fig:accum}, showing that the chosen relaxation times result in phonons along the ab-axis having MFPs on the order of microns to tens of microns, as might be expected due to the high basal plane thermal conductivity. On the other hand, phonons along the c-axis have MFPs up to 1 micron and typically in the hundreds of nanometers range, much shorter than those along the ab axis but still considerably longer than prior estimates \cite{tanaka_thermal_1972}. The calculated MFPs are, however, of the same order as those recently observed in experiment \cite{fu_experimental_2015}  and calculated with molecular dynamics. \cite{wei_phonon_2014}

\begin{figure}
\begin{center}
\includegraphics[width=\textwidth]{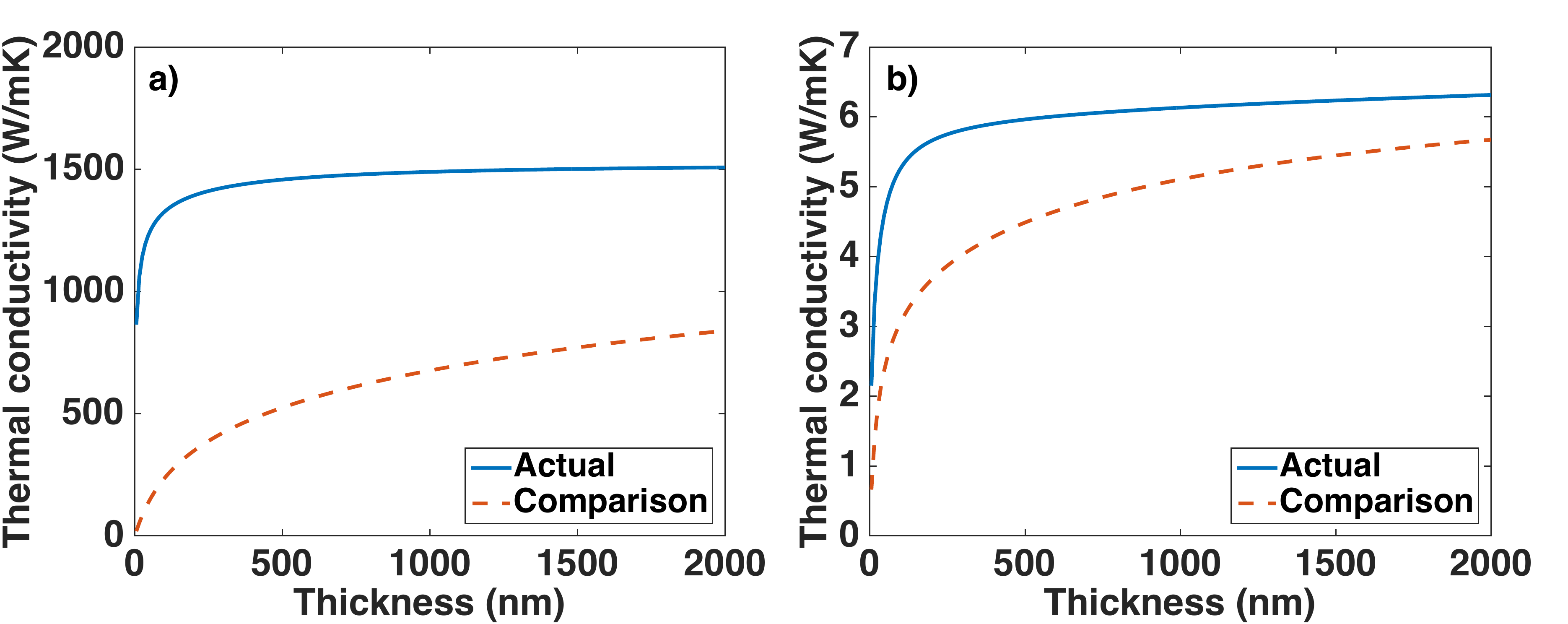}
\caption{(a) Actual ab-axis thermal conductivity using Eq. 6 (solid line) versus film thickness and comparison using the same equation with the ab-axis rather than c-axis component of the MFPs in the boundary scattering function (dashed line). The actual thermal conductivity remains high even for thicknesses as small as 10 nm due to the small MFP component along the thickness direction of the film. (b) Corresponding plot for c-axis thermal conductivity. Counterintuitively, the correct use of the ab-axis MFPs in Eq. 6 leads to a higher thermal conductivity than using the c-axis MFPs, even though the ab-axis MFPs are larger than those along the c-axis. }
\label{fig:kcalc}
\end{center}
\end{figure}

The large difference in MFP between the ab and c axes affects thermal transport in thin films because the strength of boundary scattering only depends on the component of MFP normal to the boundary. In the case of transport in a thin film along the ab-axis, the far shorter c-axis MFP leads to only minimal boundary scattering even for film thicknesses as small as 10 nm. 

The ab-axis thermal conductivity versus film thickness obtained using Eq.\ \ref{kzz} is shown in Fig.\ \ref{fig:kcalc}a. The ab-axis thermal conductivity remains close to 1000 W/mK even at thicknesses of around 10 nm, and nearly reaches its ultimate value at a thickness of around 1 micron. These observations are quite unexpected if one considers that the MFPs along the ab-axis are tens of microns as in Fig.\ \ref{fig:accum}. 

To place these calculations in perspective, we seek to calculate the thermal conductivity of an equivalent isotropic crystal with the same ab-axis MFPs. However, we were unable to find a simple way to convert the highly anisotropic graphite dispersion to an equivalent isotropic one. As a means of comparison, we instead calculate the thermal conductivity reduction using the longer ab-axis MFPs in the boundary scattering function rather than c-axis MFPs in the same figure. The result shows the trend expected of an isotropic solid, with the thermal conductivity approaching zero for very thin films and slowly approaching the bulk value as the film thickness increases. This prediction differs greatly from the correct calculation because of the form of Eq.\ \ref{kzz}: the thermal conductivity reduction depends only on the component of the MFP along the thickness direction, regardless of the ab-axis MFPs. As a result, even extremely thin graphite films maintain their high thermal conductivity despite the long MFPs along the ab-axis.

\begin{figure}
\begin{center}
\includegraphics[width=0.8\textwidth]{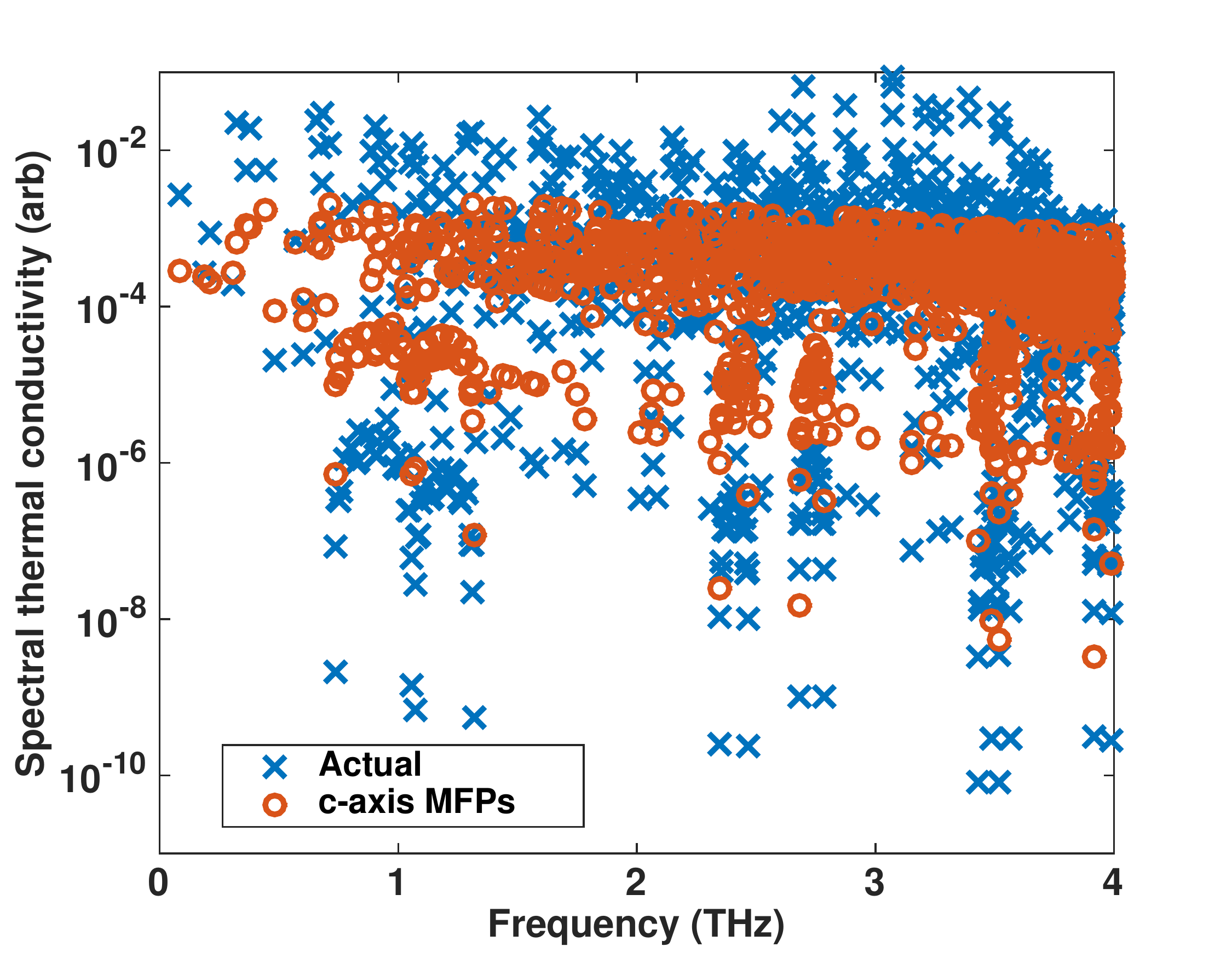}
\caption{Spectral c-axis thermal conductivity versus phonon frequency using Eq. 6 (crosses) and with the c-axis MFPs in the boundary scattering function (circles). Even though the ab-axis MFPs are longer than those in the c-axis, the correct boundary scattering function only minimally affects phonons with long c-axis MFPs, resulting in a higher thermal conductivity than if the shorter c-axis MFPs are used in the boundary scattering function.}
\label{fig:spectralk}
\end{center}
\end{figure}

The same calculation as above but for a thermal gradient along the c-axis is shown in Fig.\ \ref{fig:kcalc}b. In this case, the graphite is infinite along the c-axis and has a finite thickness along one basal axis, similar to the geometry used in a prior experimental study. \cite{fu_experimental_2015} The thermal gradient exists along the c-axis of the graphite. The thermal conductivity versus film thickness is presented according to Eq.\ \ref{kzz}, which uses the ab-axis MFPs, and for comparison the same calculation with the c-axis MFPs in the boundary scattering function. The calculation shows that, counterintuitively, the c-axis thermal conductivity obtained with the correct ab-axis MFPs in the boundary scattering function is actually higher than that with the c-axis MFPs, even though the ab-axis MFPs are considerably longer than those in the c-axis. The reason is that modes that primarily contribute to c-axis thermal conductivity have short in-plane MFPs and thus are minimally affected by the boundary function in Eq.\ \ref{kzz}. In contrast, using Eq.\ \ref{kzz} with the c-axis MFPs incorrectly reduces the contribution of the long MFP c-axis modes that contribute the most to c-axis thermal conductivity, resulting in a smaller thermal conductivity than predicted by Eq.\ \ref{kzz}. 

This point is illustrated in Fig.\ \ref{fig:spectralk}, which shows the spectral c-axis thermal conductivity $\kappa_{c,\k}=C_{\k} v_{c\k}^{2} \tk S(\xi_{y})$ versus phonon frequency calculated using the ab- and c-axis components of the MFPs in the boundary scattering function $S(\xi_{y})$. The correct boundary scattering function based on the ab-axis MFPs only minimally affects the phonons that primarily contribute to c-axis heat conduction because these phonons have short ab-axis MFPs and long c-axis MFPs. The primary effect of the correct boundary scattering function is to make the MFPs of phonons with small c-axis MFPs even smaller, thereby resulting in only a small reduction to the thermal conductivity. On the other hand, the incorrect boundary scattering function significantly reduces the MFPs of phonons with long c-axis MFPs, resulting in a anomalously large reduction in thermal conductivity. These calculations show the importance of considering the dispersion anisotropies of a crystal when considering boundary scattering in thin films.

In summary, we have derived a generalized Fuchs-Sondheimer theory for crystals with arbitrary anisotropies and applied it to thin graphite films. We find that the large velocity anisotropy in graphite causes boundary scattering to have only a minimal effect on ab-axis thermal conductivity in films as thin as 10 nm. This observation has important implications for heat-spreading in electronic devices using thin graphite films and heat conduction in graphite foams.

The author thanks Lucas Lindsay for providing the graphite dispersion and for useful discussions. This work was sponsored in part by the National Science Foundation under Grant no. CBET CAREER 1254213 and by Boeing under the Boeing-Caltech Strategic Research \& Development Relationship Agreement.

\bibliographystyle{is-unsrt}
\bibliography{Graphiteboundary}

\end{document}